\documentclass[useAMS,usenatbib,a4paper]{mn2e}
\usepackage{graphicx}
\usepackage{natbib}
\usepackage{amssymb}
\usepackage{amsmath}
\usepackage{grffile}
\usepackage{longtable,lscape}
\usepackage{color}
\usepackage{hyperref}
\newcommand{\changes}[1] {{\color{black} #1}}
\newcommand{\changestwo}[1] {{\color{black} #1}}

\newcommand{\changesfour}[1] {{\color{black} #1}}

\title[A new estimator for galaxy-matter correlations]
  {RCSLenS: A new estimator for large-scale galaxy-matter correlations}
\author[A. Buddendiek et al.]
  {A.~Buddendiek$^{1}$\thanks{abuddend@astro.uni-bonn.de},
   P.~Schneider$^{1}$,
   H.~Hildebrandt$^{1}$,
   C.~Blake$^{2}$,
   A.~Choi$^{3}$,
   T.~Erben$^{1}$,
   \newauthor
   C.~Heymans$^{3}$,
   L.~van Waerbeke$^{4}$,
   M.~Viola$^{5}$,
   J.~Harnois-Deraps$^{4}$,
   L.~Koens$^{3}$,
   \newauthor
   R.~Nakajima$\,^{1}$
   \\
  $^1$Argelander-Institut f\"{u}r Astronomie, University of Bonn, Auf dem H\"{u}gel 71, DE-53121 Bonn, Germany \\
  $^2$Centre for Astrophysics \& Supercomputing, Swinburne University of Technology, P.O. Box 218, Hawthorn, 
      VIC 3122, Australia \\
  $^3$Scottish Universities Physics Alliance, Institute for Astronomy, University of Edinburgh, Royal Observatory, 
      Blackford Hill, \\Edinburgh, EH9 3HJ, U.K. \\
  $^4$Department of Physics and Astronomy, University of British Columbia, 6224 Agricultural Road, Vancouver, 
      V6T 1Z1, B.C., Canada \\
  $^5$Leiden Observatory, Leiden University, Niels Bohrweg 2, 2333 CA Leiden, The Netherlands
 \\
  }
  
\date{\today}

\def\LaTeX{L\kern-.36em\raise.3ex\hbox{a}\kern-.15em
    T\kern-.1667em\lower.7ex\hbox{E}\kern-.125emX}

\begin{document}

\label{firstpage}

\maketitle

\begin{abstract}
\changes{We present measurements of the galaxy bias $b$ and the \changestwo{galaxy-matter} 
cross-correlation coefficient $r$ 
for the BOSS LOWZ luminous red galaxy sample.  Using a new statistical weak lensing analysis of the 
Red Sequence Cluster Lensing Survey (RCSLenS) we find the bias properties of this sample to be 
higher than previously reported with $b=2.45_{-0.05}^{+0.05}$ and $r=1.64_{-0.16}^{+0.17}$ 
\changestwo{on scales between $3\arcmin$ and $20\arcmin$}.  We repeat the measurement for angular scales of 
$20'\leq \vartheta \leq70'$, which yields $b=2.39_{-0.07}^{+0.07}$ and $r=1.24_{-0.25}^{+0.26}$.  
This is the first application of a data compression analysis using a complete sets of discrete 
estimators for galaxy-galaxy lensing and galaxy clustering.  As cosmological data sets grow,  
our new method of data compression will become increasingly important in order to interpret joint 
weak lensing and galaxy clustering measurements and to estimate the data covariance.  In future 
studies this formalism can be used as a tool to study the large-scale structure of the Universe 
to yield a precise determination of cosmological parameters.}
\end{abstract}


\section{Introduction}
\changes{Since the discovery of the accelerated expansion of the Universe 
(\citealt{riess}; \citealt{perlmutter})}
the origin and nature of dark energy remains unknown. Several possible
explanations like a cosmological constant, quintessence, or a
modification of gravity on cosmological scales have been
suggested. Although the accelerated expansion has been confirmed using a
combination of other
cosmological probes like CMB experiments (\citealt{wmap_paras};
\citealt{planck_paras}), weak gravitational lensing
(\citealt{schrabback}; \citealt{heymans}), galaxy clusters
(\citealt{vikhlinin}; \citealt{mantz}), or baryonic acoustic
oscillations (BAO, \citealt{blake}; \citealt{sanchez}) the statistical
power of these probes so far remains insufficient to reveal the true
nature of dark energy.  Statistical precision sufficient to
distinguish a cosmological constant from a more dynamical nature of
dark energy will only be reached by the next generation of cosmology
experiments, like \textit{Euclid} \citep{euclid}, the LSST \citep{lsst}, or
\textit{WFIRST} \citep{wfirst}. For this purpose the \textit{Euclid}
satellite will not only map the whole extragalactic sky in the optical
and the near-infrared, but it will also take near-infrared spectra of
about 50 million galaxies up to a redshift of $z=2$. Using this vast
data set the \textit{Euclid} consortium will measure the geometry of
the Universe using both baryonic acoustic oscillations and cosmic
shear. 

Cosmic shear is the distortion of light bundles from distant sources
caused by the intervening tidal gravitational field, caused by the
large-scale matter distribution in the Universe, which is measured
from the auto-correlation of galaxy shapes (e.g. \citealt{bacon};
\citealt{vanwaerbeke}; \citealt{hoekstra_cosmicshear}; see
\citealt{bartelmann} for a review). The gravitational lensing signal
in the galaxy shapes contributes only a few per cent to the whole
galaxy ellipticity; furthermore, these galaxies are intrinsically
small, typically smaller than the point-spread function of
ground-based observations, and correspondingly are measured \changes{over
a limited number of \changesfour{CCD} pixels.} Correcting for PSF
effects and pixelization still \changes{poses a challenge} to the
astronomical community (e.g. \citealt{great10}; \citealt{great3}).
Due to these technical difficulties, it is important to have multiple
independent weak lensing probes to map the density field in our
Universe. A particularly promising approach is to combine information
from galaxy auto-correlations (e.g.  \citealt{blake};
\citealt{sanchez}) and galaxy-matter correlations
(e.g. \citealt{edo1}; \citealt{edo2}; \citealt{velander}). 
\changes{Significant effort has been made to develop a} 
new theoretical framework for these measurements 
(e.g. \citealt{leauthaud}; \changes{\citealt{cacciato}}; \citealt{eriksen1};
\citealt{coupon}).

A further challenge in relating observed signals to theoretical
predictions stems from the difficulty \changes{in understanding baryonic physics,
such as cooling, star formation and feedback. This} affects the
statistical properties of the large-scale structure on small
scales. \changes{A particularly interesting approach was therefore}
suggested by \citet{baldauf}, henceforth B10, who introduced a new
estimator $\Upsilon$ for clustering and lensing, which \changes{eliminates} all
small-scale contributions to the signals. This \changes{methodology was successfully
applied to the} Sloan Digital Sky Survey (SDSS) by
\citet{mandelbaum}. \changes{The} new estimator can be used to constrain
cosmological parameters as well as the bias between galaxies and the
dark matter distribution.

\changes{In this work we show} that the $\Upsilon$ statistic is a special
case of the aperture mass formalism (\citealt{aperture_mass1};
\citealt{aperture_mass2}).  Using this information we generalise the
B10 approach; in particular, we define a complete set of estimators
for a given range of scales which all are `blind' to the correlation
functions below a predescribed threshold. We expect that the first few
elements of this discrete set contain all the relevant information,
which thus leads to a substantial data compression and \changes{a lower}
dimensional covariance, similar to the \changes{COSEBI-statistic} for cosmic shear
\citep{schneider}.  

As a proof of concept, in this paper we fix the cosmology and use the
new estimators to measure \changes{the galaxy bias} of a particular galaxy sample.  For
this study we use as lenses the galaxies from the Baryon Oscillation
Spectroscopic Survey (BOSS) LOWZ sample \citep{eisenstein} and as
sources photometrically selected background galaxies from the Red
Sequence Cluster Lensing Survey (RCSLenS\footnote{www.rcslens.org};
\changes{Hildebrandt et al. in prep.}).  \changes{In order to establish the accuracy of
  the estimator and to create the corresponding covariance matrix we
  use mock catalogues, that are} based on the simulations by \citet{mocks}.

\changes{Galaxy bias describes
  how galaxies trace the underlying dark matter field \citep{kaiser2}. 
  In this analysis, 
  we concentrate on the linear bias factor $b$, which is defined as
  the square root of the ratio of the galaxy and dark matter power
  spectra, and the galaxy-matter cross-correlation coefficient $r$. The bias of the
  LOWZ sample was measured in \cite{chuang} and in a different
  way by \cite{boss_lowz}, the bias of the CMASS sample for example in
  \cite{nuza_2013}. The WiggleZ sample was analysed in
  \cite{marin_2013} using 3-point correlation functions. Measuring
  these parameters is crucial for redshift-space-distortion studies as
  well as many cosmological measurements where $b$ and $r$ represent
  nuisance parameters.}

\changes{This is the first measurement of galaxy bias using
  galaxy-galaxy lensing in RCSLenS, a re-analysis of the Red Sequence
  Cluster Survey 2 (RCS2, \citealt{rcs2}). In \cite{vanuitert} a similar measurement
  has been carried out on RCS2 using correlation functions instead of
  the advanced statistics we introduce here. \cite{blake2} present galaxy-galaxy 
  measurements on the RCSLenS data to constrain modified gravity. The focus of this study is 
  to introduce a new methodology and the benefits of data
  compression taking galaxy bias measurements as an example.}

\changes{This paper is is organised as follows. In Sect.} \ref{method} we introduce the B10 method, our generalisation,
and the approach to measure the galaxy bias.  Section \ref{analysis}
describes the data analysis, and in Sect. \ref{discussion} we give a
detailed discussion.  As the fiducial cosmology we use a flat
$\Lambda$CDM cosmology constrained by \textit{Planck} with
$H_{0}=67.74\, \mathrm{km\,s^{-1}Mpc^{-1}}$,
$\Omega_{\mathrm{m}}=0.3089$, $\Omega_{\Lambda}=0.6911$, and
$\sigma_{8}=0.8159$ \citep{planck_paras}. \changes{To test} the sensitivity
of our results with respect to cosmological parameters, we also use
the cosmology obtained in \citet{heymans}: $H_{0}=73.8
\mathrm{km\,s^{-1}Mpc^{-1}}$, $\Omega_{\mathrm{m}}=0.271$,
$\Omega_{\Lambda}=0.729$, and $\sigma_{8}=0.799$. 

\section{Method}
\label{method}
\subsection{The $\Upsilon$ statistics interpreted as
  $M_{\mathrm{ap}}$} 
In B10 two new estimators were introduced, one in terms of the
projected galaxy correlation function $\omega_{\rm p}$, and one in
terms of the differential surface mass density $\Delta \Sigma$ around
galaxies\changes{. This is measured using \changes{weak gravitational lensing, namely}} 
the tangential shear component $\gamma_{\rm
  t}$. \changes{These estimators are simultaneously} analysed in order to
recover information about the dark matter distribution. In this
\changes{section} we will generalise these estimators, but instead of
$\omega_{\rm p}$ and $\Delta \Sigma$ we will use the angular
correlation function $\omega(\vartheta)$ and the tangential shear
$\gamma_{\mathrm{t}} (\vartheta)$ around (foreground) galaxies. 
These quantities can be obtained from large photometric lensing
surveys \changes{where spectroscopic redshift
information is not available.} When using only photometric redshifts,
measuring $\omega_{\rm p}$ is not sensible. 
\changes{Nevertheless, for this proof of concept study, we make use of
  a spectroscopically selected galaxy sample. This simplifies the
  interpretation of the results since the spectroscopic sample has a
  well-defined redshift and galaxy type distribution. Furthermore, it is possible 
  to measure the galaxy bias for such a sample by different means, 
  like higher-order clustering or redshift-space distortions. While measuring
  angular correlation functions for galaxies with spectroscopic
  redshifts might seem unnecessary, doing so makes this technique
  directly applicable to future photometric surveys that lack
  spectroscopy.}

The estimator introduced by B10 in case of the tangential shear
$\gamma_{\mathrm{t}}$ is\footnote{As mentioned before
  B10 actually define $\Upsilon$ in terms of $\Delta \Sigma$. To be
  consistent throughout the paper we use $\gamma_{\rm{t}}$. Thus we
  denote the B10 statistics in terms of $\gamma_{\rm{t}}$ as
  $\hat\Upsilon$. }
\begin{equation}
\label{eq:ups_mandel}
\hat\Upsilon(\vartheta, \vartheta_{\mathrm{min}}) = \gamma_{\mathrm{t}}(\vartheta)
      -\left ( \frac{\vartheta_{\mathrm{min}}}{\vartheta} \right)^{2} \gamma_{\mathrm{t}}(\vartheta_{\mathrm{min}})\;, 
\end{equation}
where $\vartheta_{\mathrm{min}}$ is the scale below which small-scale
information is suppressed. There are two features in the definition of
$\hat\Upsilon(\vartheta, \vartheta_{\mathrm{min}})$ which require
attention. First, it is a continuous function of the scale
$\vartheta>\vartheta_{\rm min}$\changes{. In any analysis, the signal needs to be measured in bins of 
$\vartheta$. This means that the angular scale needs to be
discretized when comparing measurements with theoretical predictions. 
It is usually} unclear how this discretization is optimized, as \changes{there is a balance
between having enough points} to include all relevant cosmological information
on the one hand, and to limit the number of points for \changes{a manageable
covariance matrix} on the other hand. A second feature is the
occurrence of $\gamma_{\mathrm{t}}(\vartheta_{\mathrm{min}})$ for
every $\vartheta$ in $\hat\Upsilon$, which means that any uncertainty
in this quantity will affect $\hat\Upsilon(\vartheta,
\vartheta_{\mathrm{min}})$ at all scales $\vartheta$. 
Furthermore, as the tangential shear at a fixed angular
separation cannot be measured, but must be averaged over a finite
interval, this can introduce systematics in the measurement of
$\gamma_{\mathrm{t}}(\vartheta_{\mathrm{min}})$, and thus the
$\hat\Upsilon(\vartheta, \vartheta_{\mathrm{min}})$. In fact,
\citet{mandelbaum} determined
$\gamma_{\mathrm{t}}(\vartheta_{\mathrm{min}})$ by a power-law fit of
the tangential shear (more precisely, of $\Delta\Sigma$) over a finite
interval bracketing both sides of the minimum scale.

Here we address \changes{all these} issues, by first relating the
$\hat\Upsilon$-statistic to the aperture mass \citep{aperture_mass1},
which is defined as 
\begin{equation}
 M_{\mathrm{ap}} = \int^{\phi_{\mathrm{max}}}_{\phi_{\mathrm{min}}} 
                   \mathrm d\phi \: \phi \: \mathcal{U}(\phi) \: \kappa(\phi)\;, 
\end{equation}
where $\kappa(\phi)$ is the convergence, azimuthally averaged over
polar angle and over the foreground galaxy population, 
$\mathcal{U}$ is a compensated filter function, i.e.,
\begin{equation}
 \int_{\phi_{\mathrm{min}}}^{\phi_{\mathrm{max}}} \mathrm d\phi \: 
      \phi \: \mathcal{U}(\phi) = 0 \;;
\end{equation}
and $\phi_{\mathrm{min}}$ and $\phi_{\mathrm{max}}$ 
the inner and outer scales on which the weight function is non-zero. 
The aperture mass can be expressed  
in terms of the azimuthally averaged tangential 
shear $\gamma_{\mathrm{t}}$, yielding
\begin{equation}
\label{eq:map}
 M_{\mathrm{ap}} = \int^{\phi_{\mathrm{max}}}_{\phi_{\mathrm{min}}} 
                   \mathrm d\phi \: \phi \: \mathcal{Q}(\phi) \: \gamma_{\mathrm{t}}(\phi)\; , 
\end{equation}
where $\mathcal{Q}$ is related to $\mathcal{U}$ via
\begin{equation}
\label{eq:q}
 \mathcal{Q}(\phi) = \frac{2}{\phi^{2}} \int_{0}^{\phi} \mathrm d\phi^{\prime} \: \phi^{\prime} \:
 \mathcal{U}(\phi^{\prime}) -\mathcal{U}(\phi)\;.
\end{equation}
For every value of $\vartheta$ we can interpret $\hat\Upsilon$ as an
aperture mass. Indeed, comparing Eq.\,(\ref{eq:map}) with
Eq.\,(\ref{eq:ups_mandel}), we see immediately that
$\hat\Upsilon(\vartheta, \vartheta_{\mathrm{min}})$ is a special case
of $ M_{\mathrm{ap}}$ if we set $\phi_{\rm min}=\vartheta_{\rm min}$,
$\phi_{\rm max}=\vartheta$, and 
\begin{equation}
\label{eq:ups_old_q}
 \mathcal{Q}(\phi) = + \frac{1}{\phi}\delta_{\rm D}(\phi-\vartheta) 
                     -
                     \frac{\vartheta_{\mathrm{min}}}{\vartheta^{2}}\delta_{\rm
                       D}(\phi-\vartheta_{\mathrm{min}})\;,
\end{equation}
where $\delta_{\rm D}$ is the Dirac delta function. Inverting
Eq.\,(\ref{eq:q}) we find 
\begin{equation}
 \mathcal{U}(\phi) = -\mathcal{Q}(\phi) + 2 \int_{\phi}^{\infty}\mathrm{d}\phi^{\prime}\frac{\mathcal{Q}(\phi^{\prime})}{\phi^{\prime}}\;, 
\end{equation}
which yields
\begin{align}
 \mathcal{U}(\phi) = & - \frac{1}{\phi}\delta_{\rm D}(\phi -\vartheta) 
                          +
                          \frac{\vartheta_{\mathrm{min}}}{\vartheta^{2}}\delta_{\rm
                            D}(\phi-\vartheta_{\mathrm{min}})\nonumber \\ 
                     & + \frac{2}{\vartheta^{2}} \left[ \mathcal{H}(\vartheta-\phi) 
                      - \mathcal{H}(\vartheta_{\mathrm{min}}-\phi) \right]\;,
\end{align}
where $\mathcal{H}$ is the Heaviside step function. This equation
shows that the $\hat\Upsilon$-statistics is indeed insensitive to
$\kappa(\vartheta)$ on scales $\vartheta<\vartheta_{\rm min}$, and
thus allows the exclusion of small scales where theoretical
\changes{predictions are currently uncertain.}

\subsection{Measuring $\Upsilon$ by using a set of orthogonal
  functions} 
The filter functions $\mathcal{U}$ and $\mathcal{Q}$ of the aperture
mass depend on the scale $\vartheta$ of $\hat\Upsilon$. Instead of
using a continuum of scales $\vartheta$, we can define a complete set
of compensated filter functions $\mathcal{U}_n$ over the range of
scales $\vartheta_{\rm min}\le\vartheta\le\vartheta_{\rm max}$, i.e., each filter
function satisfies
\begin{equation}
 \int_{\vartheta_{\mathrm{min}}}^{\vartheta_{\mathrm{max}}} \mathrm d\vartheta \: 
      \vartheta \: \mathcal{U}_{n}(\vartheta) = 0 \;.
\end{equation}
The completeness ensures that the corresponding set of aperture masses
contains the full information contained in $\hat\Upsilon(\vartheta,
\vartheta_{\mathrm{min}})$ for $\vartheta_{\rm
  min}\le\vartheta\le\vartheta_{\rm max}$. In fact, we expect that
most of the information is \changes{included} in only the first few elements of this
set, whereas the remaining ones contain essentially only
noise. \changesfour{This is
due to the fact that the weight functions ${\mathcal U}_n$ are ordered
according to their number of roots, together with the fact that the
galaxy-galaxy lensing signal is not expected to contain substantial
small-scale structure.} Working with a few numbers, instead of a continuous function,
will ease the analysis, in particular the generation of covariances,
due to the associated data compression, while keeping the essential
features of $\hat\Upsilon$, i.e., suppression of small-scale
influence. 

\changes{Given the many other studies measuring galaxy bias for
  BOSS galaxies it is clear that the data compression is not
  crucial for this kind of measurement. However, with future surveys
  becoming increasingly large and the desire to split the huge galaxy
  samples into many sub-samples (in redshift, type, etc.) it will
  become more important to minimise the size of the data vector. Since
  mock catalogues need to be used to estimate covariances their required number
  directly scales with the number of elements in the data vector. This
  study represents a simple test case that can be directly compared to
  the literature in order to validate the method.}

\begin{figure}
 \centering
 \includegraphics[width=8.5cm,height=7cm,keepaspectratio=true]{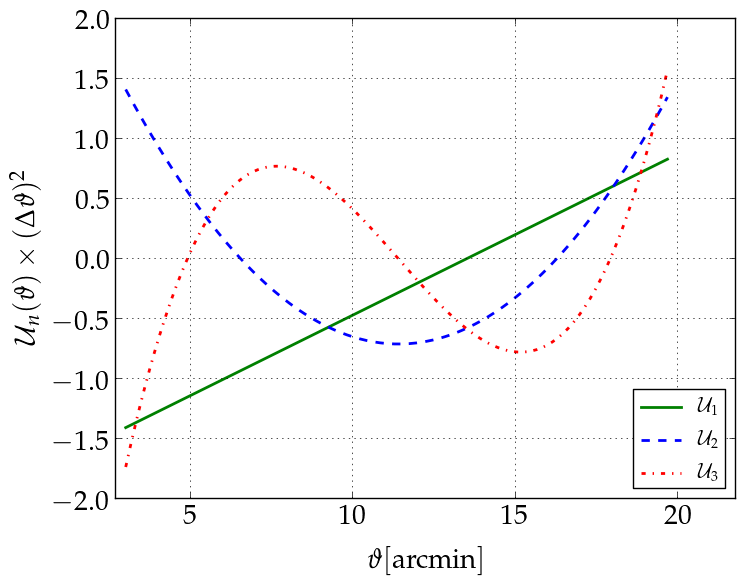}
 \caption{Filter functions $\mathcal{U}_{n}(\vartheta)$ for clustering 
          \changes{defined by Eq. (\ref{eq:u_filter1}) and Eq. (\ref{eq:u_filter2})} 
          for $\vartheta_{\mathrm{min}}=3\,\rm{arcmin}$ and $\vartheta_{\mathrm{max}}=20\,\rm{arcmin}$. 
          \changes{These enter the clustering estimator via Eq. (\ref{eq:ups_gg}).}}
 \label{fig:filt_clus}
\end{figure}

\begin{figure}
 \centering
 \includegraphics[width=8.5cm,height=7cm,keepaspectratio=true]{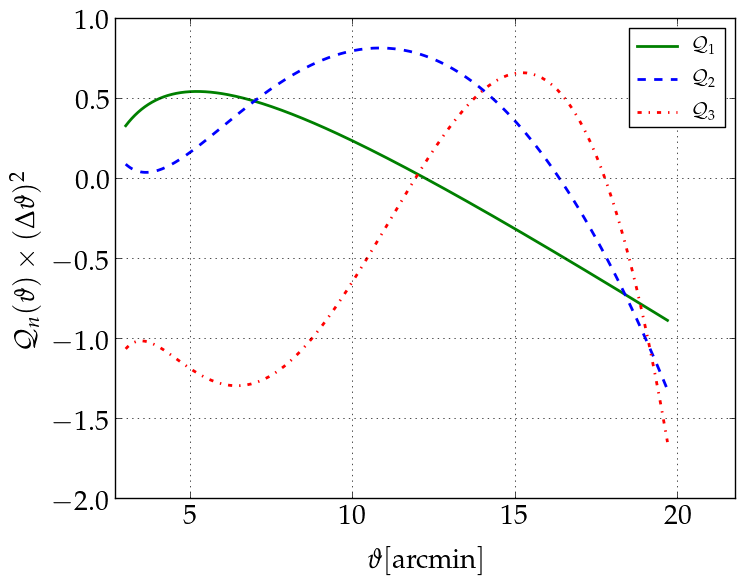}
 \caption{Filter functions $\mathcal{Q}_{n}(\vartheta)$ for lensing 
          \changes{defined by Eq. (\ref{eq:qn})} 
          for $\vartheta_{\mathrm{min}}=3\,\rm{arcmin}$ and $\vartheta_{\mathrm{max}}=20\,\rm{arcmin}$. 
          \changes{These enter the galaxy-galaxy lensing estimator via Eq. (\ref{eq:ups_gm}).}}
 \label{fig:filt_lens}
\end{figure}

We choose the filter functions to be orthogonal, i.e., 
\begin{equation}
 \int_{\vartheta_{\mathrm{min}}}^{\vartheta_{\mathrm{max}}} \mathrm d\vartheta \: 
      \mathcal{U}_{n}(\vartheta) \: \mathcal{U}_{m}(\vartheta) = 0
      \quad\hbox{for} \quad m\ne n \;. 
\end{equation}
The Legendre polynomials $\mathcal{P}_{n}$ form a complete orthogonal
set of functions on $[-1,1]$, which we can use to find a set of
suitable filter functions. 
\changes{We decide to use the Legendre Polynomials as they already have many of the desired properties 
for the filter functions.} 
For this to work we define the
transformation used in \citet{schneider}
\begin{equation}
 x=\frac{2 (\vartheta - \bar{\vartheta})}{\Delta \vartheta}\;, 
\end{equation}
with $\Delta \vartheta =
\vartheta_{\mathrm{max}}-\vartheta_{\mathrm{min}}$,
$\bar{\vartheta}=(\vartheta_{\mathrm{min}}+\vartheta_{\mathrm{max}})/2$
and $\mathrm{d} \vartheta=\frac{\Delta \vartheta}{2}\mathrm{d} x$.
This maps the interval
$[\vartheta_{\mathrm{min}},\vartheta_{\mathrm{max}}]$ onto
$[-1,1]$. Setting 
\begin{equation}
\mathcal{U}_n(\vartheta)={1 \over (\Delta\vartheta)^{2}}
u_n\left( {2 (\vartheta - \bar{\vartheta})\over
    \Delta\vartheta}\right) \;,
\label{eq:utoU}
\end{equation}
\changes{where we explicitly impose the dependence on $x$ and normalise by $1/(\Delta\vartheta)^{2}$, 
so that the $\mathcal{U}_n$ have correct units. 
This} transforms
the compensation and orthogonality conditions into
\begin{equation}
 \int_{-1}^{1}\mathrm{d}x \: \left(\frac{x \Delta \vartheta}{2}
+\bar \vartheta \right) \: u_{n}(x)=0
\end{equation}
and
\begin{equation}
\label{eq:cond2}
 \int_{-1}^{1}\mathrm{d}x \: u_{n}(x) \: u_{m}(x) = \delta_{nm}\;,
\end{equation}
where in the latter case we fixed the normalization of the filter
functions. The Legendre polynomials can be defined via the recurrence
relation
\begin{equation}
 \mathcal{P}_{n+1}(x) = \frac{1}{n+1}\left [ (2 n+1) \: x \: \mathcal{P}_{n}(x) - n \: \mathcal{P}_{n-1}(x)\right ]\;, 
\end{equation}
with $\mathcal{P}_{0}(x)=1$ and $\mathcal{P}_{1}(x)=x$. 
\changes{We first
try to find dimensionless filters $u_{n}(x)$ which are proportional to
the $\mathcal{P}_{n}(x)$}; 
these can then be transformed into the
$\mathcal{U}_{n}(\vartheta)$ according to Eq.\,(\ref{eq:utoU}).
The first function to 
fulfil our two conditions is a first-order polynomial of the form
$u_{1}(x)=a_1 x+a_0$, where the two coefficients $a_i$ are determined
from the two conditions, to yield
\begin{equation}
 u_{1}(x) = \frac{3Gx-1}{\sqrt{2(1+3G^{2})}}\;, 
\end{equation}
where we defined $G=2\bar{\vartheta}/\Delta\vartheta$. 
Since 
\begin{equation}
\label{eq:legend_prop}
 \int_{-1}^{1} \mathrm{d}x \: \mathcal{P}_{n}(x) \: x^{m} = 0
\end{equation}
for $m < n$ and because the Legendre polynomials are orthogonal we can
choose for $n\ge2$ the filter functions
\begin{equation}
 u_{n}(x) = \sqrt{\frac{2n+1}{2}} \: \mathcal{P}_{n}(x) \mathcal{H}(1-x^{2})\;, 
\end{equation}
which has the correct normalization, and we explicitly included the
finite interval of support for the $u_n$. Using Eq.\,(\ref{eq:utoU}),
we then find
\begin{align}
\label{eq:u_filter1}
\mathcal{U}_{n}(\vartheta) = & \frac{1}{(\Delta \vartheta)^{2}} u_{n}(x) \nonumber\\ 
                           = & \frac{1}{(\Delta \vartheta)^{2}} \sqrt{\frac{2n+1}{2}} 
                             \mathcal{P}_{n}\left(\frac{2 (\vartheta - \bar{\vartheta})}
                               {\Delta \vartheta}\right) \\ \nonumber
              & \times \mathcal{H}(\vartheta-\vartheta_{\mathrm{min}}) \mathcal{H}(\vartheta_{\mathrm{max}}-\vartheta)\;,
\end{align}
for $n\geq2$ and 
\begin{align}
\label{eq:u_filter2}
\mathcal{U}_{1}(\vartheta) = &\frac{1}{(\Delta\vartheta)^2}\frac{3G\left(\frac{2 (\vartheta - \bar{\vartheta})}
                                                           {\Delta \vartheta}
                                                           \right)-1}{\sqrt{2(1+3G^{2})}} \nonumber\\ 
                      & \times \mathcal{H}(\vartheta-\vartheta_{\mathrm{min}})\mathcal{H}(\vartheta_{\mathrm{max}}-\vartheta)\;.
\end{align}
The $\mathcal{Q}_{n}(\vartheta)$ follow immediately as
\begin{equation}
\label{eq:qn}
 \mathcal{Q}_{n}(\vartheta)=\frac{2}{\vartheta^{2}}\int_{0}^{\vartheta}\mathrm d \vartheta^{\prime} \; \vartheta^{\prime} \;
 \mathcal{U}_{n}(\vartheta^{\prime}) - \mathcal{U}_{n}(\vartheta)\;.
\end{equation}
The final estimators for galaxy-galaxy lensing then become
\begin{equation}
\label{eq:ups_gm}
 \Upsilon_{\mathrm{gm}}(n) = \int_{\vartheta_{\mathrm{min}}}^{\vartheta_{\mathrm{max}}} \mathrm d \vartheta \:
 \vartheta \: \: \mathcal{Q}_{n}(\vartheta) \: \gamma_{\rm t}(\vartheta) \;.
\end{equation}
\changes{We want to compare the clustering of galaxies with the
galaxy-galaxy lensing signal, to learn about the biasing of galaxies
and the \changes{cross-correlation coefficient} between the galaxies and the
underlying matter distribution. Thus, }we define integrals of the galaxy
angular correlation function that have the same angular dependence as
the filter functions for the convergence $\kappa$, i.e.,
\begin{equation}
\label{eq:ups_gg}
 \Upsilon_{\mathrm{gg}}(n) = \int_{\vartheta_{\mathrm{min}}}^{\vartheta_{\mathrm{max}}} \mathrm d \vartheta \:
 \vartheta \: \: \mathcal{U}_{n}(\vartheta) \: \omega(\vartheta)\;.
\end{equation}
Note that the clustering signal will be measured using the lens sample
from galaxy-galaxy lensing in order to probe the same density field.
During our analysis we will make use of only the first three orders of
the filter functions; for our dataset, those should contain all
relevant information. The
corresponding filter functions for
$\vartheta_{\mathrm{min}}=3\,\rm{arcmin}$ and
$\vartheta_{\mathrm{max}}=20\,\rm{arcmin}$ are displayed in
Fig. \ref{fig:filt_clus} and Fig. \ref{fig:filt_lens}.

\subsection{Connecting observables to theory}

In order to constrain cosmological parameters or to measure the bias
factor, we need to know how the observables
$\Upsilon_{\mathrm{ij}}(n)$ are connected to predictable
theoretical quantities like the three dimensional dark matter power
spectrum $\mathcal{P}_{3D}(k,w)$, where $k$ is the comoving wavenumber
and $w$ the comoving distance, characterizing the cosmic epoch. This
is now shown for the case where the lens sample has a rather broad
redshift distribution\changes{,} as for increasingly small
distributions the following approximation\changes{s for the angular correlations diverge and are not valid
any more}.

\changes{In the following we assume that the bias is linear and can be
described by
\begin{equation}
  \hat{b}^2=\frac{\mathcal{P}_{\rm gg}}{\mathcal{P}_{\rm 3D}}\,,
\end{equation}
with $\mathcal{P}_{\rm gg}(k,w)$ being the galaxy power spectrum. This
assumption is valid on large scales which we explicitly limit
ourselves to with the $\Upsilon$ formalism. Furthermore, 
we define the cross-correlation coefficient
\begin{equation}
 \hat{r} = \frac{\mathcal{P}_{\mathrm{gm}}}{\sqrt{\mathcal{P}_{\rm gg} \mathcal{P}_{\rm 3D}}}\, ,
\end{equation}
}
where $\mathcal{P}_{\mathrm{gm}}(k,w)$ is the cross-power spectrum
between matter and galaxies. $\hat{r}$ is important for determining the 
galaxy-matter cross-correlations. 

The angular correlation function of galaxies is related to
$\mathcal{P}_{3D}$ through \citep{henk_bias}
\begin{align}
\label{eq:ang_cor0}
 \omega(\vartheta) = & \frac{1}{2\pi}\int \mathrm{d}w \; \left
   (\frac{p_{\mathrm{l}w}(w)}{f_{k}(w)}\right)^{2} \nonumber \\ \times 
    & \int \mathrm{d}\ell \; \ell\; \hat{b}^{2}(\ell,z)\mathcal{P}_{3D}\left(\frac{\ell}{f_{k}(w)};w\right) \mathrm{J_{0}}(\ell\vartheta)\;, 
\end{align}
where $\hat{b}(\ell,z)$ is the galaxy bias as a function of angular
wave number $\ell=k\,f_k(w)$ and redshift $z$,
$w$ the \changes{comoving} distance,
$f_{k}(w)$ the comoving angular diameter distance,
$p_{\mathrm{l}w}(w)$ the lens probability distribution in terms of $w$, and
$\mathrm{J_{0}}$ the zeroth-order Bessel function of the first kind.
Changing the order of
integration and replacing the probability distribution with respect to
$w$, $p_{\mathrm{l}w}(w)$, by the observable redshift distribution, using
\mbox{$\;p_{\mathrm{l}z}(z)\mathrm{d}z=p_{\mathrm{l}w}(w)\mathrm{d}w$}, 
yields
\begin{align}
\label{eq:ang_cor}
 \omega(\vartheta) = & \frac{1}{2\pi}\int \mathrm{d}\ell \; \ell\;
 \mathrm{J_{0}}(\ell\vartheta)  \\ \nonumber \times
    & \int \mathrm{d}w \left (\frac{p_{\mathrm{l}z}(z)}{f_{k}(w)}\right)^{2} 
    \left(\frac{\mathrm{d}z}{\mathrm{d}w}\right)^{2} \hat{b}^{2}(\ell,z)\mathcal{P}_{3D}\left(\frac{\ell}{f_{k}(w)};w\right)\;, 
\end{align}
with 
\begin{equation} \nonumber
\frac{\mathrm{d}z}{\mathrm{d}w}=\frac{H_{0} \sqrt{(1+z)^{2}(1+z\Omega_{\mathrm{m}})-z(2+z)\Omega_{\Lambda}}}{c}\;.
\end{equation}
By inserting Eq.\,(\ref{eq:ang_cor}) into Eq.\,(\ref{eq:ups_gg}) we
obtain an expression for $\Upsilon_{\mathrm{gg}}(n)$, which depends
quadratically on the galaxy bias
\begin{align}
\label{eq:ups_gg_theory}
 & \Upsilon_{\mathrm{gg}}(n) = \frac{b^{2}}{2\pi} \int_{\vartheta_{\mathrm{min}}}^{\vartheta_{\mathrm{max}}} \mathrm d \vartheta \:
 \vartheta \: \: \mathcal{U}_{n}(\vartheta)  \\ \nonumber 
 & \times \int \mathrm{d}\ell \; \ell\; \mathrm{J_{0}}(\ell\vartheta)
    \int \mathrm{d}w \left (\frac{p_{\mathrm{l}z}(z)}{f_{k}(w)}\right)^{2} 
    \left(\frac{\mathrm{d}z}{\mathrm{d}w}\right)^{2} \mathcal{P}_{3D}\left(\frac{\ell}{f_{k}(w)};w\right)\;.
\end{align}
Here we defined $b$ as a weighted average of the bias
$\hat{b}(\ell,z)$ over $\ell$ and $z$, where the weight is given by
the factors in the \changes{second integral} in \changes{Eq. (\ref{eq:ang_cor})}. We point
out that $b$ still depends on the order $n$ (due to the dependence of
the angular weight function $\mathcal{U}_n$ on $\vartheta$), which we
do not write out explicitly\footnote{When constraining $b$ later on,
  we will actually constrain an average over $n$, $\ell$, and $z$}.
The connection between $\mathcal{P}_{3D}$ and
$\gamma_{\mathrm{t}}(\vartheta)$ has been shown to be
(\citealt{kaiser}; \citealt{guzik})
\begin{align}
 \label{eq:tan_shear}
 \gamma_{\mathrm{t}}(\vartheta) = & \frac{3\; \Omega_{\mathrm{m}}}{4\pi}\left(\frac{H_{0}}{c} \right)^{2} 
        \int \mathrm{d}w \; \frac{g(w)p_{\mathrm{l}w}(w)}{a(w) f_{k}(w)}  \\ \nonumber
        & \times \int \mathrm{d}\ell\; \ell\; \hat{b}(\ell,z)\, \hat{r}(\ell,z)
        \mathcal{P}_{3D}\left(\frac{\ell}{f_{k}(w)};w\right)
        \mathrm{J_{2}}(\ell\vartheta) \;,
\end{align}
where $\hat{r}$ is the cross-correlation coefficient, $a(w)$ the cosmic scale factor, and $g(w)$ is the
mean of angular \changes{diameter} distances \changes{\citep[e.g.,][]{aperture_mass2}}
\begin{equation}
g(w) = \int_{w}^{w_{H}}\mathrm{d}w^{\prime}\;p_{\mathrm{s}w}(w^{\prime}) \frac{f_{k}(w^{\prime}-w)}{f_{k}(w^{\prime})}\;, 
\end{equation}
where $p_{\mathrm{s}w}(w)$ is the source distance probability
distribution in terms of $w$.  Again, by changing the order of
integration, inserting the redshift probability distribution and inserting it
into Eq.\,(\ref{eq:ups_gm}), one finds
\begin{align}
\label{eq:ups_gm_theory}
 & \Upsilon_{\mathrm{gm}}(n) = \frac{3\; \Omega_{\mathrm{m}}}{4\pi}\left(\frac{H_{0}}{c} \right)^{2} b\, r 
    \int_{\vartheta_{\mathrm{min}}}^{\vartheta_{\mathrm{max}}} \mathrm d \vartheta \:
 \vartheta \: \: \mathcal{Q}_{n}(\vartheta) \\ \nonumber
 & \times \int \mathrm{d}\ell\; \ell\; \mathrm{J_{2}}(\ell\vartheta)
   \int \mathrm{d}w \; \frac{g(w)p_{\mathrm{l}z}(z)}{a(w) f_{k}(w)} \frac{\mathrm{d}z}{\mathrm{d}w}\mathcal{P}_{3D}\left(\frac{\ell}{f_{k}(w)};w\right)\; . 
\end{align}
As before, we use the weighted average of $\hat{b}$ and $\hat{r}$ over $\ell$,
$z$ and $\vartheta$. 
When measuring $\Upsilon_{\mathrm{gm}}(n)$ and $\Upsilon_{\mathrm{gg}}(n)$ from the data, we can simultaneously fit the models 
to both signals. In this way we can either 
\begin{enumerate}
 \item fix the cosmology and constrain $b$ and $r$,
 \item fix $b$ and $r$ and constrain the cosmology, 
 \item \changes{set $r=1$ and fit $b$ and the cosmology simultaneously,}
 \item or constrain $b$, $r$, and the cosmology simultaneously. 
\end{enumerate}
The latter is possible by combining galaxy clustering and
galaxy-galaxy lensing with a cosmic shear signal, weighted by the same
kernel functions $\mathcal{U}_n(\vartheta)$. Since the scope of this
work is to proof the concept we will use a fixed cosmology and
constrain the galaxy bias $b$ and the cross-correlation coefficient
$r$.

\section{Data Analysis}
\label{analysis}

\changes{We choose to apply our new methodology to determine a large-scale bias measurement
of the BOSS LOWZ sample. This sample is well suited for this first analysis, as there are already 
measurements and it is less complicated compared to a whole cosmological 
study.} 

\subsection{Data sets}

\subsubsection{BOSS LOWZ}
\begin{figure}
 \centering
 \includegraphics[width=8.5cm,height=7cm,keepaspectratio=true]{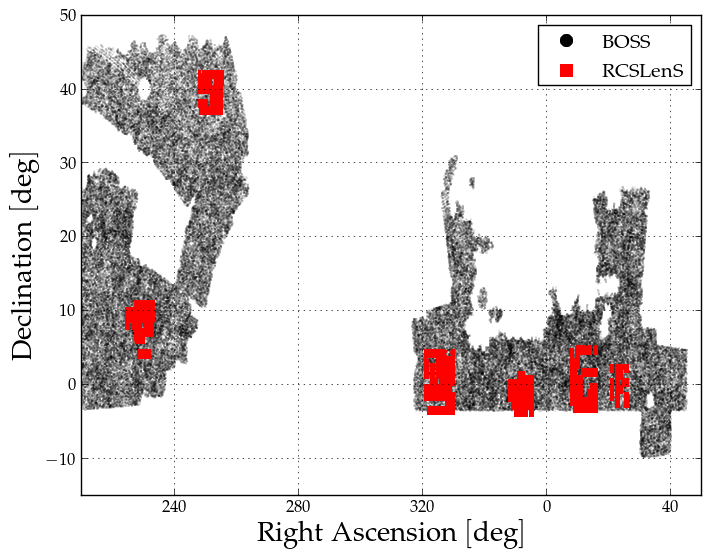}
 \caption{RCSLenS \changes{regions} that were used, and the galaxies from BOSS. The RCSLenS \changes{regions} are 
 non-contiguous 
 because of the lack of four-band data, which is needed for photometric redshifts. }
 \label{fig:patches}
\end{figure}

\changes{We measure the weak lensing signal around galaxies from BOSS  \citep{eisenstein}, using}
the 10th Data Release \citep{ahn}. 
\changes{We select galaxies following \citet{chuang} and \citet{sanchez} to select a spectroscopic 
redshift sample with  $0.15\leq z \leq 0.43$. This yields 9102
galaxies within the RCSLenS footprint.} 
For the lensing measurements we
only use the BOSS galaxies that lie within the BOSS-RCSLenS overlap;
however, for the clustering measurement, the whole LOWZ
sample is used, which is spread over a much larger area \changes{($\sim 5\,000\, \mathrm{deg^{2}}$, \citealt{tojeiro})} 
and consists of $218\,891$ galaxies. 
In this way we can make use of the much better statistics arising from the larger sample. 
\changes{This is a \changestwo{valid} approach as in Section \ref{sec:results} we show that the 
signals measured for both samples are consistent with each other.} 
The BOSS and RCSLenS overlapping area is shown in Fig. \ref{fig:patches}. 
The summed $p_{\mathrm{l}z}(z)$ derived from
spectroscopic redshifts of the lenses can be seen in
Fig. \ref{fig:pofz}.  For the clustering measurements we make use of
the weights, $\Theta$, provided by the BOSS collaboration, which
account for fiber collisions as explained in \citet{anderson}.

\subsubsection{RCSLenS}
\begin{figure}
 \centering
 \includegraphics[width=8.5cm,height=7cm,keepaspectratio=true]{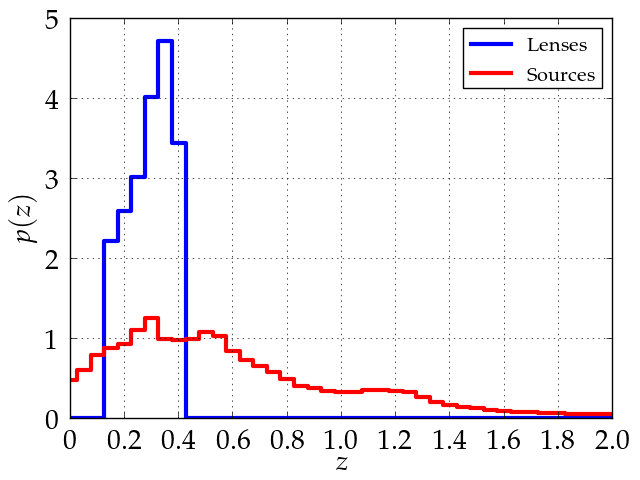}
 \caption{$p_{\mathrm{l}z}(z)$ of lenses (blue) and
   $p_{\mathrm{s}z}(z)$ of sources (red). For the lenses we use the spectroscopic redshifts 
 to estimate $p_{\mathrm{l}z}(z)$, whereas for the sources we make use of the stacked full $p(z)$ of every source 
 galaxy, which is estimated by the photometric redshift code. \changes{The distributions are normalised so that 
 $\sum p(z) \Delta z = 1$.} \changestwo{Additionally, we weight the distributions using the weights described 
 in Section \ref{analysis}}}
 \label{fig:pofz}
\end{figure}

RCSLenS \changes{(Hildebrandt et al., in prep.)} is an analysis of the original
\changes{RCS2} using the
\changes{Canada France Hawaii Telescope Lensing Survey pipeline (CFHTLenS;} 
\citealt{hildebrandt}; \citealt{heymans_cfht};
\citealt{miller}; \citealt{erben}) \changes{to reduce the data and create}
shape and photometry catalogues.  The survey was carried out
using Megacam at the \changes{Canada France Hawaii Telescope (CFHT)} 
and has only one exposure per band per pointing.
It covers roughly 500 $\mathrm{deg}^{2}$ in the $g^{\prime}$-,
$r^{\prime}$-, $i^{\prime}$- and $z^{\prime}$-band and \changes{with} an additional
250 $\rm{deg}^{2}$ with three or fewer \changes{bands}.  The
$r^{\prime}$-band is used as the lensing band with a $5\sigma$ point
source limiting magnitude of $m_{\mathrm{lim}}=24.3$ and a median
seeing of $0.71\, \mathrm{arcsec}$ \citep{rcs2}. Galaxy shapes are
measured using \emph{lens}fit \citep{miller}. As described in
\citet{blake2} we use the \emph{lens}fit weights $\eta$ and the BOSS
weights $\Theta$ for the lensing analysis. We take both weights in
order to use the same weighting scheme in the lensing as well as in
the clustering analysis. 
\changesfour{The resulting estimator is
\begin{equation}
 \langle \gamma_{\mathrm{t}} (\vartheta)\rangle = 
    \frac{\sum_{i, \mathrm{sources}}\sum_{j, \mathrm{lenses}} \eta_{i} \Theta_{j} e_{\mathrm{t},i,j}}
	 {\sum_{i, \mathrm{sources}}\sum_{j, \mathrm{lenses}} \eta_{i} \Theta_{j}}, 
\end{equation}
Here $\eta_{i}$ denotes the \emph{lens}fit weight of the $i$th
source galaxy and $\Theta_{j}$ the BOSS weight of the $j$th lens
galaxy, whereas $e_{\mathrm{t},i,j}$ is the tangential ellipticity of the $i$th source with respect to 
the $j$th lens.}
For selecting source
galaxies, we only use the six \mbox{RCSLenS} \changes{regions} that have four
band photometry and sufficient overlap with BOSS. Those are CDE0133,
CDE0047, CDE1645, CDE2329, CDE1514, and CDE2143. 
\changes{This leaves us with about $170\, \rm{deg^{2}}$ in area and $4\,657\,415$ source galaxies.}
\changes{As sources we select all galaxies with a \changes{\emph{lens}fit} weight $\eta>0$ that 
are outside of masks.} 
\changes{We use the posterior redshift distribution for each source galaxy, estimated 
with the photometric redshift code \texttt{BPZ} \citep{benitez}, to find the 
summed $p_{\mathrm{s}z}(z)$ of the sources, which is displayed in Fig. \ref{fig:pofz}.}

\begin{figure}
 \centering
 \includegraphics[width=8cm,height=7cm,keepaspectratio=true]{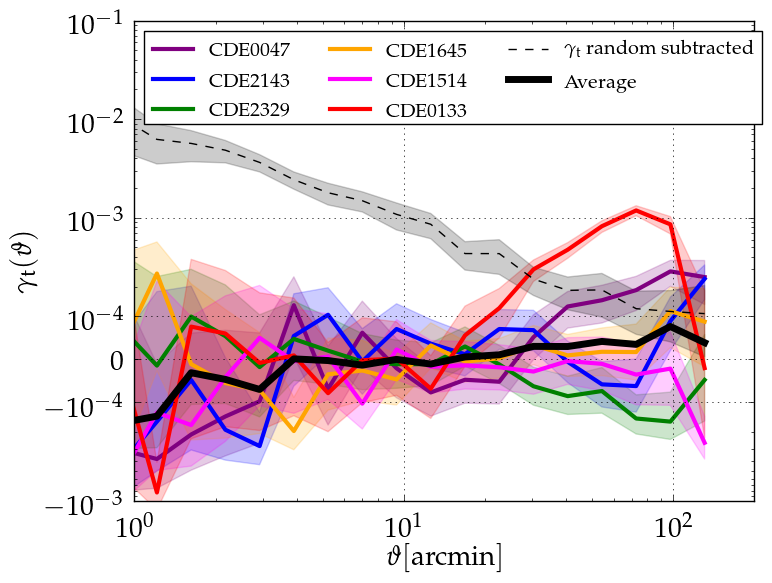}
 \caption{The lensing signal around random points. \changesfour{The coloured lines show the signal 
          for every \changes{region}, whereas the black solid line shows the average. Furthermore, 
          we display the measured signal of $\gamma_{\mathrm{t}}$ around BOSS LOWZ galaxies as 
          the dashed black line. The shaded regions correspond to the 1$\sigma$ errors.} 
          The strongest random signal corresponds to CDE0133, which is the 
          smallest \changes{region} in the area we use, and thus it has the smallest impact on the 
          total signal. \changes{For the measurements we subtract the signal for each region separately.}
          }
 \label{fig:random}
\end{figure}

\changes{The shear}
measurements for RCSLenS suffer from a multiplicative as well as an
additive bias so that
\begin{equation}
\changes{\langle e_{\mathrm{obs}} \rangle = (1+\langle m\rangle ) \langle e_{\mathrm{true}} \rangle + c\;,  }
\end{equation}
as explained for example in \citet{miller}. 
Here $e_{\mathrm{obs}}$ is the observed ellipticity of a galaxy image, $e_{\mathrm{true}}$ the 
\changes{sheared} intrinsic ellipticity, $1+m$ the correction factor for the 
multiplicative bias ($m$-correction), and $c$ is the correction for the 
additive bias ($c$-correction). 
We correct the measured
shapes of galaxies for the multiplicative bias using the factor $(1+m)$
determined for every galaxy (for more details see
e.g. \changes{\citealt{miller}} or \changes{Hildebrandt et al., in prep.}).  We apply the
$m$ correction as an ensemble correction in order to avoid
correlations between the correction and the intrinsic shape of the
galaxy \citep{miller}
\begin{equation}
 \langle \gamma^{\mathrm{cal}}_{\mathrm{t}}(\vartheta) \rangle 
          = \frac{\langle \gamma_{\mathrm{t}}(\vartheta)\rangle}{1+K(\vartheta)}\;,
\end{equation}
where 
\begin{equation}
 1+K(\vartheta) = \frac{\sum \eta_{i}\Theta_{j}(1+m_{i})}{\sum \eta_{i}\Theta_{j}}\;. 
\end{equation}
\changesfour{As before,} $\eta_{i}$ denotes the \emph{lens}fit weight of the $i$th
source galaxy and $\Theta_{j}$ the BOSS weight of the $j$th lens
galaxy. \changes{The sums are taken over all lens-source pairs separated by the angle $\vartheta$. 
The correction $1+K(\vartheta)$ is of the order 0.95
for all scales used.} \changesfour{As common in galaxy-galaxy lensing studies (e.g., 
\citealt{mandelbaum2006}), we do not apply an additive $c$-correction
but subtract the $\gamma_{\mathrm{t}}$ signal measured around random
points\changes{, which is equivalent to a direct $c$-correction for galaxy-galaxy-lensing
measurements}.} 
\changes{To determine this correction the} number of random points used depends on 
the \changes{region} size and differs between
$\sim100,000$ and $\sim180,000$. The measured signal around random points 
 is consistent with zero on scales below
$30-40\,\mathrm{arcmin}$ and rises out to larger scales, where for
\changes{$\vartheta>40\,\mathrm{arcmin}$ it can reach an amplitude of a few
times $10^{-4}$ for some \changes{regions}.} We subtract this signal for every
\changes{region} separately as it would average out when combined from all
\changes{regions}. The signals are shown in Fig. \ref{fig:random}. The \changes{region} with the strongest 
random signal is CDE0133, which is the smallest in area and thus contributes the least to the total signal. 

For the weighted average source density we find \changes{\mbox{$\sim
   5.1 \, \mathrm{galaxies}/\mathrm{arcmin}^{2}$}} when using
\begin{equation}
 n_{\mathrm{eff}} = \frac{1}{A_{\mathrm{eff}}}\frac{\left( \sum \eta_{i}\right)^{2}}
                                                   {\sum (\eta_{i})^{2}}\;,
\end{equation}
as defined in \citet{heymans_cfht}, where
\changes{\mbox{$A_{\mathrm{eff}}=174.32\,\mathrm{deg^{2}}$}} is the total
unmasked area \changestwo{in the BOSS-RCSLenS overlap}. We use this definition to account for the fact that we
use the \emph{lens}fit weight in the analysis.  The RCSLenS
catalogues are also subject to a blinding scheme. In order to avoid
confirmation bias the galaxy ellipticities exist in four versions A,
B, C, and D. One of them is the true measured one, whereas the rest
have been changed by a small factor as described in \changes{Hildebrandt et al. (in prep.) for RCSLenS
and in \citet{kuijken} for the KiDS survey}. This analysis has been performed
four times using the different ellipticity versions. After the
analysis had been finished the \changes{lead author} contacted the 
external blinder, \changes{Matthias Bartelmann,} who revealed which catalogue was the \changes{truth}.  
We then used the results of the true measured ellipticities only. 
No changes were \changes{made} after ``unblinding''.
For more information
about \mbox{RCSLenS} and the data production process we refer to
\changes{Hildebrandt et al. (in prep.)}. 

\subsection{Mock catalogues}
In order to \changes{estimate} the covariance of the $\Upsilon$s, we make use of the
simulations described in \citet{mocks}.  Those have box sizes of
$505\,h^{-1}\rm{Mpc}$, $1536^{3}$ particles each and are on $3072^{3}$
grids, which are projected onto $12288^2$ pixels. The light cones are
then extracted from those onto $6000^{2}$ pixels grids. 
The cosmology used is $\Omega_{\rm{m}}=0.2905$,
$\Omega_{\Lambda}=0.7095$, $\sigma_{8}=0.826$, and $H_{0}=68.98\,
\mathrm{km\,s^{-1}Mpc^{-1}}$.  The slight difference to the cosmologies we use
will introduce a small systematic error in the covariance, \changes{which we will neglect 
in this study.}

Based on these simulations we use a set of mock catalogues 
\changestwo{designed}
to match the properties of the
RCSLenS sources and the BOSS LOWZ lenses. They specifically match the
ellipticity and redshift distributions of RCSLenS and the clustering
properties of the LOWZ sample. 
\changesfour{We apply photometric redshift scatter to the mock sources through a $z_{\rm{spec}}$-$z_{\rm{phot}}$ matrix 
calibrated from the BPZ redshift probability distributions. The mock LOWZ lenses are added to the simulation using a halo 
occupation distribution approach calibrated by matching the observed clustering amplitude.}
\changes{In total we use 360 mock catalogues, 
which are $60\,\rm{deg^{2}}$ each}. \changesfour{The size of the region used 
for the mocks is just determined by the size of the simulations themselves. We do not aim to simulate the whole 
survey area, but for practicality we area-scale the covariance from the $60\,\rm{deg^{2}}$ outputs.}
Using six of the mocks we can create one mock survey, assuming
that each of the six RCSLenS \changes{regions} fits within the
$60\,\rm{deg^{2}}$. \changes{This then results in 60 mock realisations of RCSLenS.} 
Whenever the \changes{regions} are too big we use as much
area as possible and scale the covariance accordingly
by using the ratio of the area of the mock \changes{region} and the real \changes{region}. 
Furthermore,
for the covariance estimation \changes{we use only} the
BOSS-RCSLenS overlap for the measurements of the clustering signal,
whereas for the data we use the whole BOSS area. In order to account
for this we rescale the clustering part of the covariance with the
ratio of the two areas. \changesfour{Additionally, we set the cross-covariance 
between $\Upsilon_{\rm{gg}}$ and $\Upsilon_{\rm{gm}}$
to $0$}, as the BOSS-RCSLenS overlap is just a small fraction of the
whole BOSS area. This has been shown to be a valid approach by
\citet{more}, who conduct similar measurements with BOSS and the
CFHTLenS catalogues.  In the end we have 60 mock surveys, to which we
apply the same masks as for the data set. \changes{For this we neglect that the
mocks assume a flat sky, as the resulting differences are clearly negligible 
compared to the statistical error of our measurements given the small 
\changes{extent} of each \changes{region}.}

\subsection{Measuring two-point correlations}
\begin{figure}
 \centering
 \includegraphics[width=8.5cm,height=7cm,keepaspectratio=true]{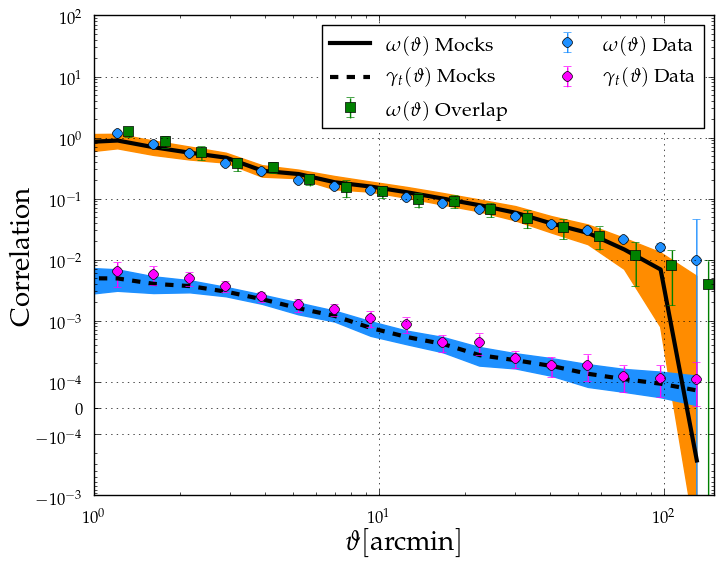}
 \caption{\changes{Comparison of galaxy clustering and galaxy-galaxy lensing signals in the mocks and data}. 
          The black lines show the mean; 
          \changes{the $1\sigma$ standard deviation} is indicated by the blue and yellow shaded regions.  
          The measurement from the data is shown as the blue and magenta points. They are in good agreement with the 
          mocks. Additionaly, the clustering signal measured just for the BOSS-RCSLenS overlap is displayed as the 
          green points. This is consistent with the signal from the whole LOWZ sample. }
 \label{fig:mocks}
\end{figure}

Before we can determine the \changes{compressed observables} $\Upsilon_{\mathrm{ij}}(n)$, we first need
to measure the corresponding galaxy-galaxy lensing and galaxy
clustering signals. \changes{We choose to measure these} in two intervals
\begin{enumerate}
 \item $3'\le \vartheta \le 20'$,
 \item $20'\le\vartheta \le 70'$
\end{enumerate}
in 200 linear bins. \changes{The \changestwo{centre of the first range corresponds to comoving length of 
$\sim3\,\mathrm{Mpc}$ at a redshift of 
$z\approx0.29$, the second one to a comoving length of $\sim12\,\mathrm{Mpc}$.} These are both large-scale, which 
will enables us to measure the large-scales bias of the LOWZ sample.} 
As a cross-check we also determine these signals
for a larger angular scale in larger logarithmic bins. The 200 linear
bins will later be used for determining the $\Upsilon$.  For
$\omega(\vartheta)$ we use the Landy-Szalay estimator \citep{landy}. 
\changes{We show the mean signals for $\gamma_{\mathrm{t}}$ and $\omega$ measured in the mocks
together with the real data in Fig. \ref{fig:mocks}. Those measurements are in good agreement. }

\subsection{$\Upsilon_{\mathrm{gm}}(n)$ and $\Upsilon_{\mathrm{gg}}(n)$}
\label{sec:results}
\begin{figure}
 \centering
 \includegraphics[width=8cm,height=8cm,keepaspectratio=true]{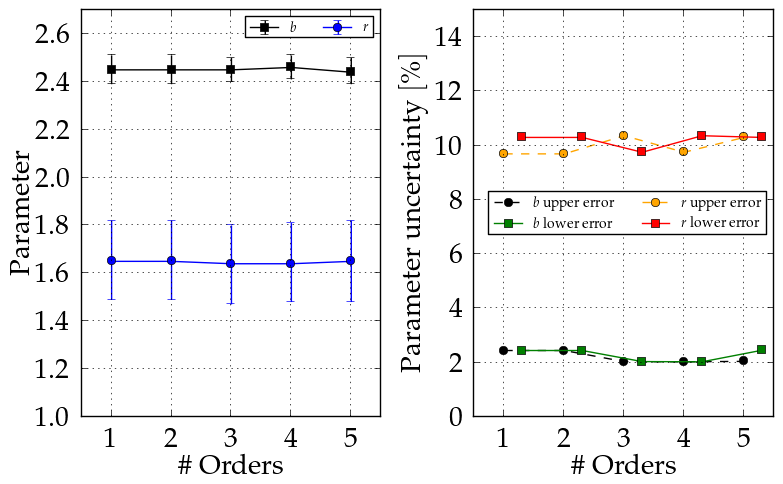}
 \caption{\changes{\textit{Left:} The measured parameter values as
	  \changestwo{a function of maximum $\Upsilon$-order $n$ for $b$ and $r$.}
          No significant difference in the values is visible, from which we conclude that the data compression is 
          indeed working and only a few orders contain all the information from the measured signals. 
          \textit{Right:} The parameter uncertainty in per cent for $b$ and $r$, again as a 
          function of $n$. Here, we do not find a significant difference, which again shows that 
          the data compression of $\Upsilon(n)$ is robust.}
          }
 \label{fig:order_info}
\end{figure}

\begin{figure*}
 \centering
 \includegraphics[width=17cm,height=10cm,keepaspectratio=true]{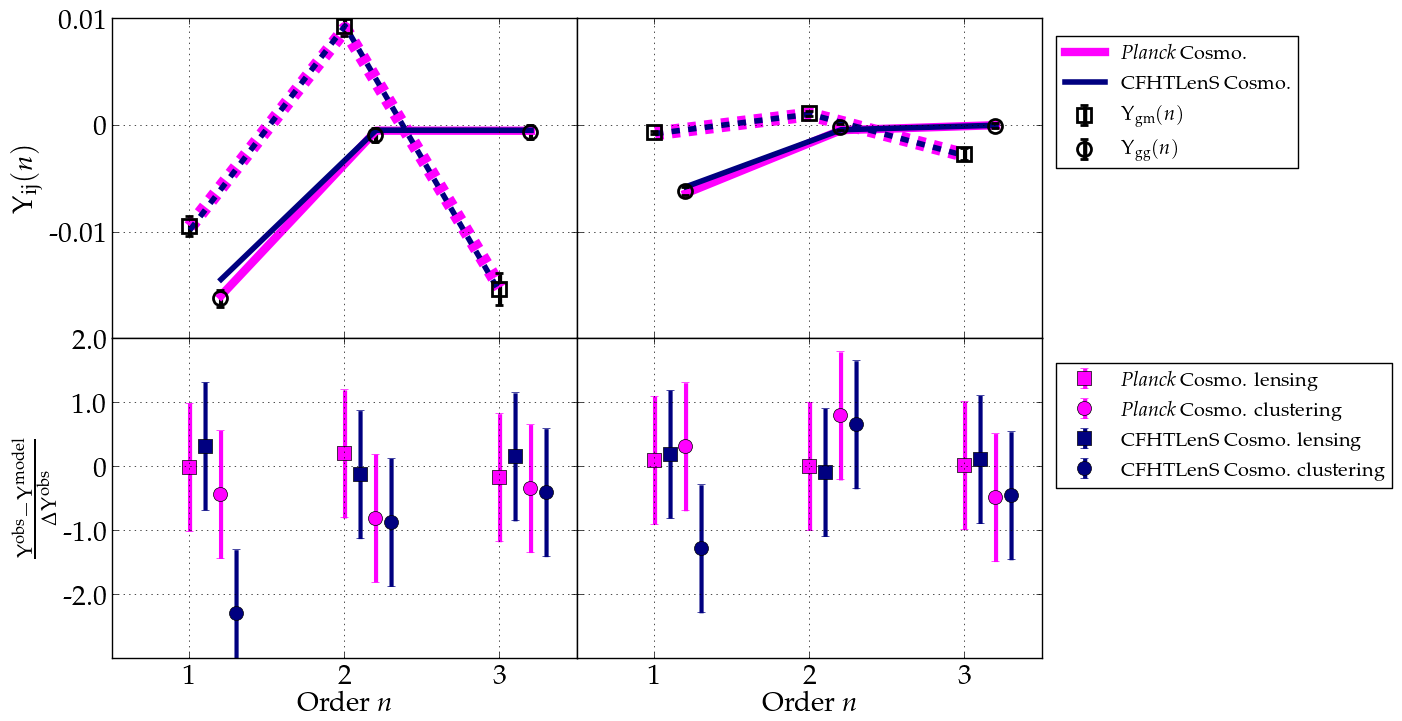}
 \caption{\changes{The top panels show the measured $\Upsilon_{\mathrm{gm}}$ and 
          $\Upsilon_{\mathrm{gg}}$ and 
          the best fit using one of the two cosmologies}. \changesfour{The magenta and dark blue 
          lines are the connections between the predicted data points using the \textit{Planck} or the 
          CFHTLenS cosmology}. In the bottom panels we show the residuals 
          $(\Upsilon^{\mathrm{obs}}-\Upsilon^{\mathrm{model}})/\Delta \Upsilon^{\mathrm{obs}}$
          \changestwo{, where $\Delta \Upsilon^{\mathrm{obs}}$ is the uncertainty in the measured 
          $\Upsilon$}.  
          \textit{Left:} Measurements for the $3-20\,\rm{arcmin}$ interval. \changes{Clearly, on these scales 
          the model we adopt to describe the galaxy bias is not a good description of the data 
          shown here, especially the clustering data. For more details see Section \ref{sec:results}}
          \textit{Right:} Measurements for the $20-70\,\rm{arcmin}$ interval. 
          }
 \label{fig:upsilons}
\end{figure*}

\begin{figure}
 \centering
 \includegraphics[width=8.5cm,height=6cm,keepaspectratio=true]{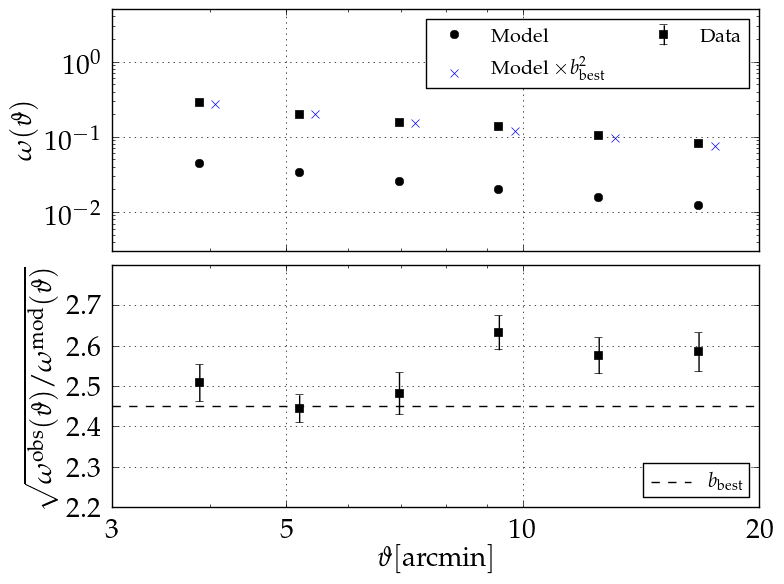}
 \caption{\changes{\textit{Top:} The angular correlation function within the 
          small scales interval, the corresponding model for $b=1$, and the best fit model. \changesfour{This best
          fit model has been determined from a joint fit of $\Upsilon_{\rm{gm}}$ and $\Upsilon_{\rm{gg}}$.}
          \textit{Bottom:} The square root of the ratio between the measured $\omega(\vartheta)$ 
          and the model one, which is an estimator for $b$. Apparently, in contradiction 
          to our assumption, there is a scale dependence of $b$. This is why the data shown in the 
          left panel of Fig. \ref{fig:upsilons} is not well described by the model. A variation of 
          $b$ of about 5 per cent within this interval would already be enough to reconcile the data 
          and the model.} \changesfour{The data shown here corresponds to the \textit{Planck} cosmology
          measurements.}}
 \label{fig:complicated_bias}
\end{figure}

\begin{figure}
 \centering
 \includegraphics[width=8.5cm,height=6cm,keepaspectratio=true]{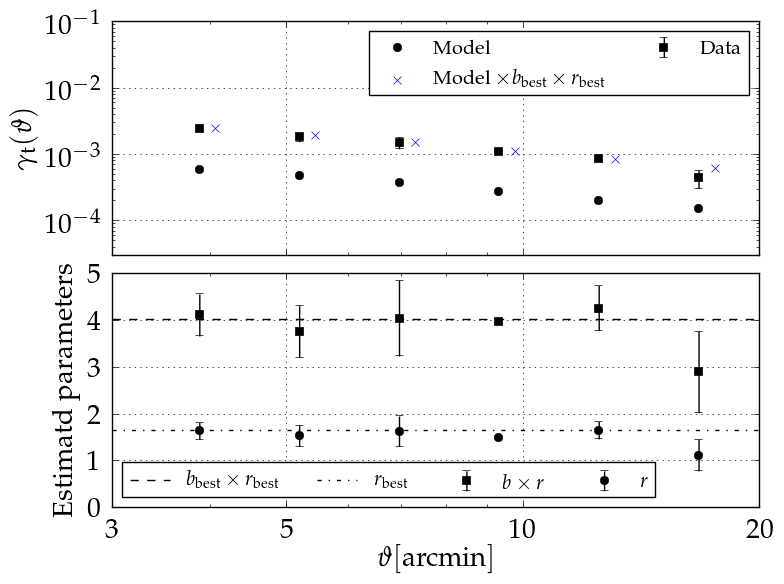}
 \caption{\changesfour{\textit{Top:} The tangential shear function within the 
          small scales interval, the corresponding model for $b,r=1$, and the best fit model. This best
          fit model has been determined from a joint fit of $\Upsilon_{\rm{gm}}$ and $\Upsilon_{\rm{gg}}$.
          \textit{Bottom:} The ratio between the measured $\gamma_{\mathrm{t}}(\vartheta)$ 
          and the model one, which is an estimator for $b\times r$. Due to the larger uncertainties 
          the shear measurements do not show a preference for scale dependent bias. Additionally, we also
          show the corresponding estimate for $r$, if we use $b$ as estimated in Fig. \ref{fig:complicated_bias}. 
          The data shown here corresponds to the \textit{Planck} cosmology measurements.}}
 \label{fig:complicated_bias_shear}
\end{figure}

\begin{figure*}
 \centering
 \includegraphics[width=16cm,height=21cm,keepaspectratio=true]{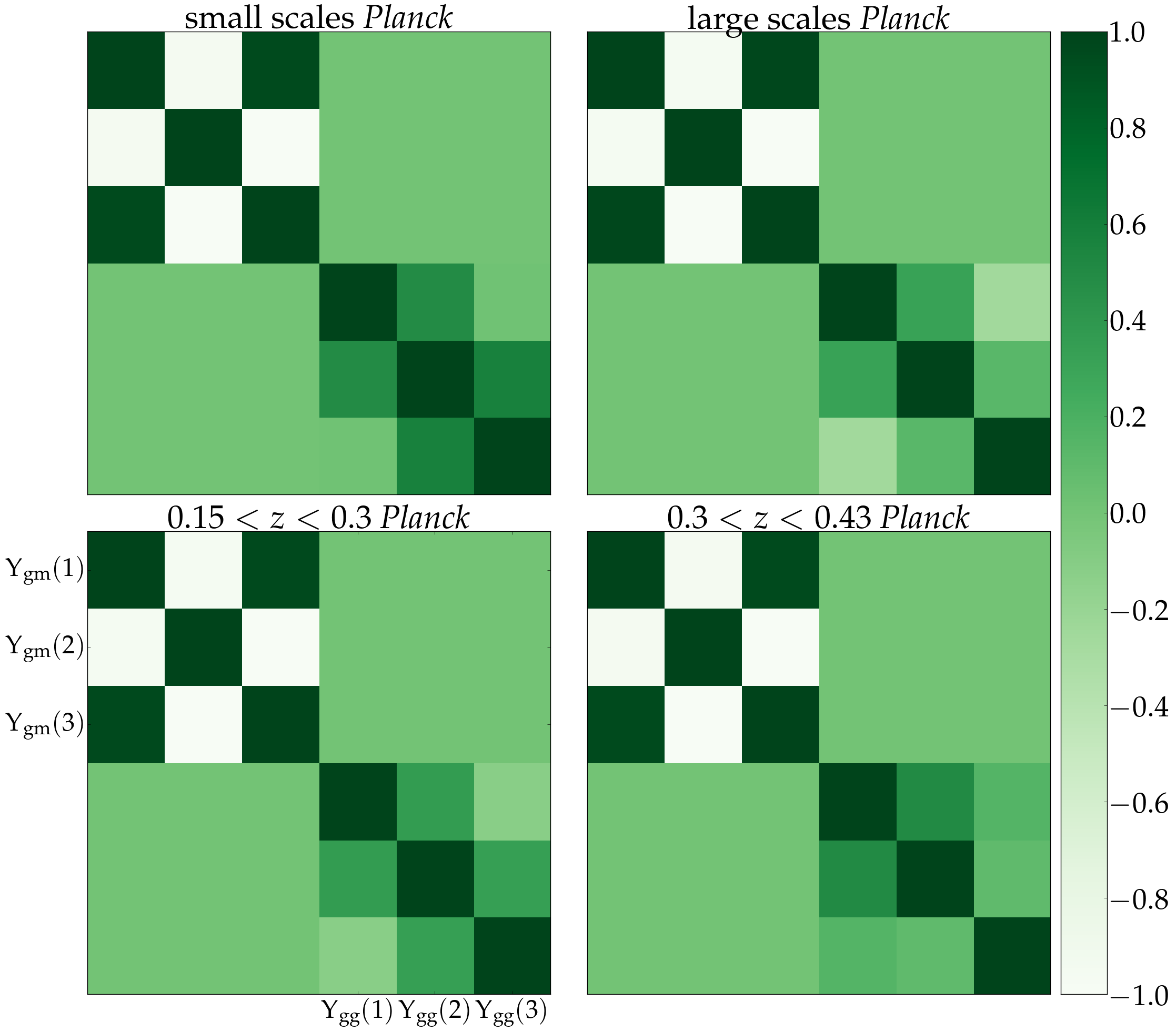}
 \caption{The correlation matrices for $\Upsilon_{\mathrm{gm}}$ and $\Upsilon_{\mathrm{gg}}$ for 
          \changes{the measurements using the \textit{Planck} cosmology. 
          \changestwo{As they only depend on the cosmology used in the mocks,}
          we do not show the correlation matrices for the CFHTLenS cosmology measurements.}
          The \changes{upper} left part of the matrices corresponds to galaxy-galaxy lensing, the bottom right to galaxy 
          clustering. \changes{The cross-covariance terms are set to $0$ as the area for the lensing measurement 
          is only a small fraction of the clustering area, which makes those measurements independent.} 
          In the order left to right, top to bottom we show the matrix for 
          the $3-20\,\rm{arcmin}$ interval and the \textit{Planck} cosmology, 
          the $20-70\,\rm{arcmin}$ interval and the \textit{Planck} cosmology, 
          the $3-20\,\rm{arcmin}$ interval and the $0.15<z<0.3$ lens sample, 
          and the $\changes{3}-20\,\rm{arcmin}$ interval and the
          $0.3<z<0.45$ lens sample. }
 \label{fig:covariance}
\end{figure*}

We use $\gamma_{\mathrm{t}}(\vartheta)$ and $\omega(\vartheta)$
measured in the 200 linear bins and integrate them using \changes{Eqs.
  (\ref{eq:ups_gm}) and (\ref{eq:ups_gg})} in order to find
$\Upsilon_{\mathrm{gm}}(n)$ and $\Upsilon_{\mathrm{gg}}(n)$. Here we
only compute the first three orders. At the end of our analysis we
tested how the parameter constraints on $b$ and $r$ changed with the
number of $\Upsilon$ orders used. We found no significant difference
for up to 5 orders and decided to use 3 orders, which yields a
sufficient number of data points for our analysis and still benefits
from a low-dimensional covariance. The fact that we do not find a
decrease of parameter uncertainty with increasing number of orders
shows that the first few orders indeed contain all the relevant
information\changes{ (see
  Fig. \ref{fig:order_info} for more details)}. The measured data
points for both angular intervals are presented in
Fig. \ref{fig:upsilons}. \changes{Unlike correlation function
  measurements these $\Upsilon$ data points cannot be interpreted
  easily. However, it is clear that in the large scales interval 
  the model (see   Section \ref{method}) is a very good fit to the data 
  regardless of the cosmological parameters used.\footnote{\changes{Note that the data
      points are highly covariant (see also
      Fig.~\ref{fig:covariance}).}} This is not the case for the smaller scales interval, 
  where \changesfour{one of the clustering data points is} several $\sigma$ away from the best fit model. 
  Clearly, the assumption of linear bias on these non-linear scales is not valid for the clustering 
  data. This is partly due to the small uncertainties in these measurements as well as the fact that 
  we neglect the model uncertainties. As can be seen 
  in Fig. \ref{fig:complicated_bias}, a change of $b$ of about 5 per cent within this interval would 
  already be enough to reconcile the data with the model.} \changesfour{If model uncertainties had been included 
  in this figure, it is likely that data and model would be in line again. Another possible explanation of 
  the discrepancy between data and model in this case 
  could be that the fiducial cosmology is wrong. Furthermore, we also investigate if an indication
  for a scale-dependent bias can be found in the $\gamma_{\mathrm{t}}(\vartheta)$ data in 
  Fig. \ref{fig:complicated_bias_shear} and find no such preference.}

From the 60 mock realizations we compute the $\Upsilon_{\mathrm{gm}}$
and $\Upsilon_{\mathrm{gg}}$ covariance matrix by measuring the
signals for each mock survey. \changes{For the inverse covariance we
take into account the correction factor from \citet{hartlap}, which
prevents us from underestimating the uncertainty in the parameter
estimates.}  The correlation matrices for all measurements are shown
in Fig. \ref{fig:covariance}. The covariance matrix is then used for a
maximum likelihood analysis, in which we simultaneously fit
theoretical predictions to $\Upsilon_{\mathrm{gm}}$ and
$\Upsilon_{\mathrm{gg}}$ with the galaxy bias $b$ and the
cross-correlation coefficient $r$ as free parameters. We compute the
predictions from \changes{Eqs. (\ref{eq:ups_gg_theory}) and
  (\ref{eq:ups_gm_theory})} using the 3D matter power spectrum
computed with \changes{\texttt{nicaea}} \citep{nicea}, which uses the
recipe from \citet{smith2}.  The resulting likelihood contours are
displayed in Fig. \ref{fig:contours}. We perform this fit twice using
the \textit{Planck} cosmology as well as \changes{the best fit cosmology} from
\changes{CFHTLenS}, constrained in \citet{heymans}, to test for the
dependence of the parameters on different cosmologies. The results are
presented in Table \ref{tab:res}.  For the maximum likelihood analysis
we assume a Gaussian likelihood function. \changes{Note that one 
cannot directly interpret the $\chi^2/{\rm dof}$ values since the model
is non-linear and the data noisy \citep{andrae_2010}. 
We find $b=2.45_{-0.05}^{+0.05}$ and $r=1.64_{-0.16}^{+0.17}$ for the
small scales interval, and for angular scales of 
$20'\leq \vartheta \leq70'$ we find $b=2.39_{-0.07}^{+0.07}$ and $r=1.24_{-0.25}^{+0.26}$.}
\begin{table}
\caption{Parameter estimates for galaxy bias $b$ and cross-correlation coefficient $r$.
         In case of the full sample the second column indicates the cosmology used. 
         For the samples used in Section \ref{sec:sanity}, it indicates which subsample was used.} 
\begin{center}
\begin{tabular}{c|c|c|c|c}
\hline \hline
scale &  & $b$ & $r$ & $\chi^{2}/\mathrm{dof}$ \\ \hline
$3\arcmin-20\arcmin$ & \textit{Planck} & $2.45^{+0.05}_{-0.05}$ & $1.64^{+0.17}_{-0.16}$ & 0.38 \\ \hline
$3\arcmin-20\arcmin$ & CFHTLenS & $2.33^{+0.05}_{-0.05}$ & $1.78^{+0.18}_{-0.18}$ & 0.53 \\ \hline
$20\arcmin-70\arcmin$ & \textit{Planck} & $2.39^{+0.07}_{-0.07}$ & $1.24^{+0.26}_{-0.25}$ & 0.47  \\ \hline
$20\arcmin-70\arcmin$ & CFHTLenS & $2.27^{+0.07}_{-0.07}$ & $1.33^{+0.28}_{-0.27}$ & 0.38 \\ \hline
$3\arcmin-20\arcmin$ & $0.15<z<0.3$ & $2.35^{+0.04}_{-0.05}$ & $1.84^{+0.24}_{-0.23}$ & 2.01  \\ \hline
$3\arcmin-20\arcmin$ & $0.3<z<0.45$ & $2.61^{+0.07}_{-0.08}$ & $1.33^{+0.21}_{-0.20}$ & 0.73 \\ \hline
\end{tabular}
\end{center}
\label{tab:res}
\end{table}

The estimated values for $b$ are slightly higher compared to the
findings by \citet{boss_lowz}, who determine the bias by fitting their
projected clustering signal to HOD populated $N$-body
simulations. Using their best fit model and the corresponding
simulations they predict the bias for the LOWZ sample as a function of
physical scale.  For $3\,\mathrm{Mpc}$, which corresponds to about
$11\,\mathrm{arcmin}$ at a redshift of $0.29$, \changes{they find} a
bias of $\sim2.2$, whereas for $12\,\mathrm{Mpc}$
($\sim45\,\mathrm{arcmin}$) it corresponds to a bias of $\sim2.1$. 
\changes{This differs by $\sim 10$ per cent from our results.}
The discrepancy \changes{could be} explained by our approach of
averaging over $\ell$ and $z$ and the corresponding weight functions
\changes{, but as there are no
error bars in \citet{boss_lowz}, we cannot judge how significant the difference is.
\citet{chuang} also measure the bias for the LOWZ sample, 
finding a value of $b\times \sigma_{8}=1.102\pm0.039$ for scales between 
$24$ and $200h^{-1}\,\mathrm{Mpc}$. This corresponds to a significantly smaller value of $b$ compared 
to the findings in this study. However, the two approaches of measuring the bias, as well as 
the scales used are very different. This discrepancy could therefore be resolved 
if we considered scale-dependent bias.}
\changestwo{Whereas one might have expected that the cross-correlation coefficient $r$ is
close to unity on these scales, we instead find
it to be $3 \sigma$} away from unity. On large scales, however, we find $r$ to be
close to unity \changes{as expected for deterministic large-scale
bias}. One should note that a measured $r>1$ is possible, as was
discussed in B10 and also found by \citet{marian} in the Millenium simulations, as the
angular galaxy correlation function is a shot-noise subtracted estimator. 
Furthermore, we point out that the values measured for different cosmologies
differ by a few percent which is smaller than the parameter
uncertainties from statistical errors.

\subsection{\changes{Redshift evolution test:} splitting up the LOWZ sample}
\label{sec:sanity}
\begin{figure*}
 \centering
 \includegraphics[width=8cm,height=7cm,keepaspectratio=true]{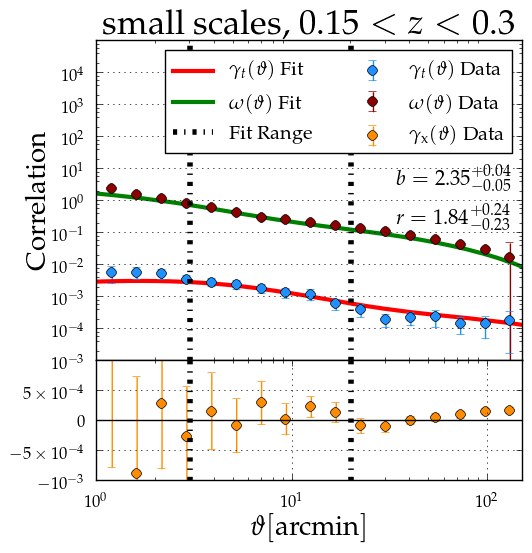}
 \includegraphics[width=8cm,height=7cm,keepaspectratio=true]{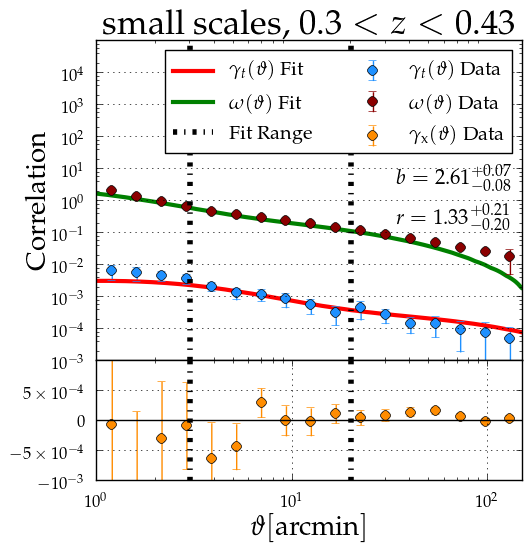}
 \includegraphics[width=8cm,height=7cm,keepaspectratio=true]{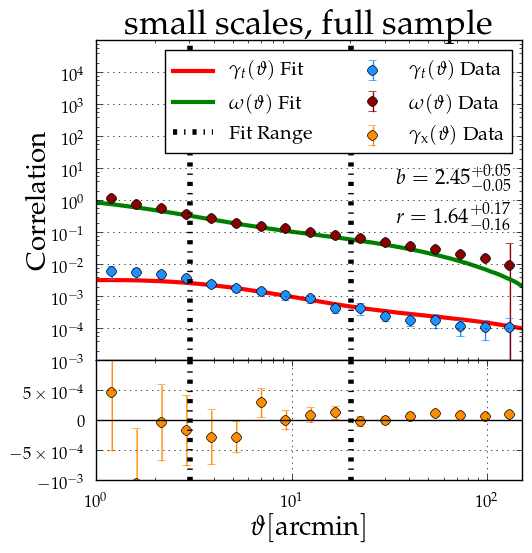}
 \caption{Galaxy clustering and galaxy-galaxy lensing signals \changes{with the best fit theoretical model 
          for the $3-20\, \rm{arcmin}$ 
          interval using a \textit{Planck} cosmology. The two sub-samples from the redshift evolution test are used 
          as well as the full sample.} The best fit lines 
          were fitted to the $\Upsilon$s, not the signals shown here. Within the fitting range the estimated parameter 
          values for $b$ and $r$ appear to be in excellent agreement with the data. 
          \changes{We also show $\gamma_{\mathrm{x}}$, which is consistent with zero in all cases. 
          A non-zero $\gamma_{\mathrm{x}}$ points to systematic issues in the data.}}
 \label{fig:corr_fit}
\end{figure*}

In Fig. \ref{fig:corr_fit} we show the measured signals for $\gamma_{\mathrm{t}}(\vartheta)$ 
and $\omega(\vartheta)$ for the whole sample as well as for the two sub-samples (described below). 
We also scale the expected signals for both with the constrained values 
of $b$ and $r$. The data is consistent with constant values of $b$ and $r$, and 
the values for both parameters obtained from the fit to the $\Upsilon$s is consistent with 
the signals of the correlation functions $\gamma_{\mathrm{t}}(\vartheta)$ 
and $\omega(\vartheta)$. \changes{This means that the method introduced here is indeed 
capable of compressing the data while not losing information contained in the correlation
functions.}

\begin{figure*}
 \centering
 \includegraphics[width=8cm,height=7cm,keepaspectratio=true]{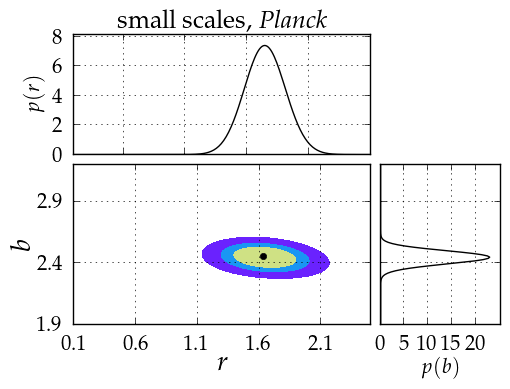}
 \includegraphics[width=8cm,height=7cm,keepaspectratio=true]{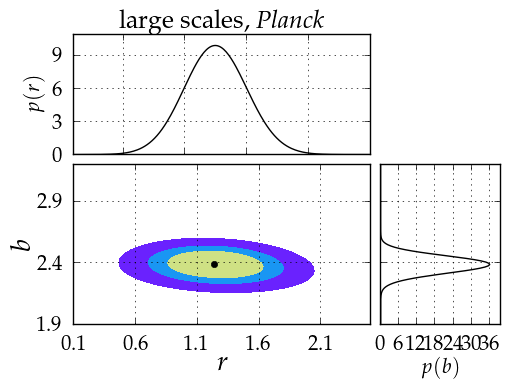}
 \includegraphics[width=8cm,height=7cm,keepaspectratio=true]{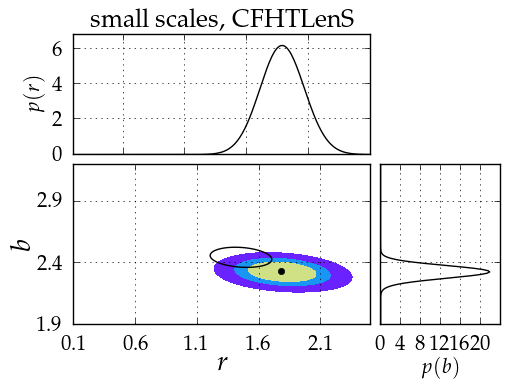}
 \includegraphics[width=8cm,height=7cm,keepaspectratio=true]{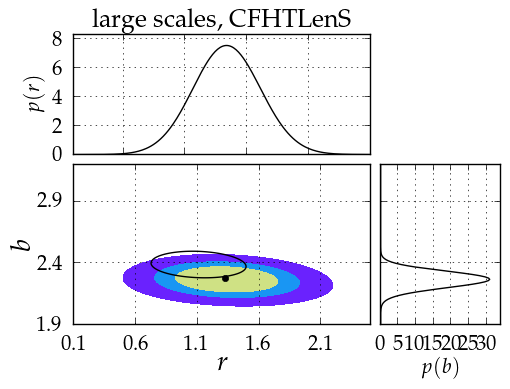}
 \includegraphics[width=8cm,height=7cm,keepaspectratio=true]{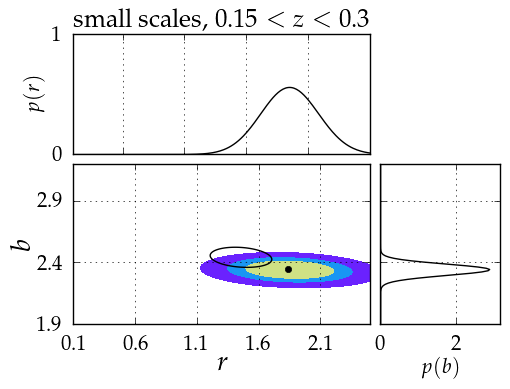}
 \includegraphics[width=8cm,height=7cm,keepaspectratio=true]{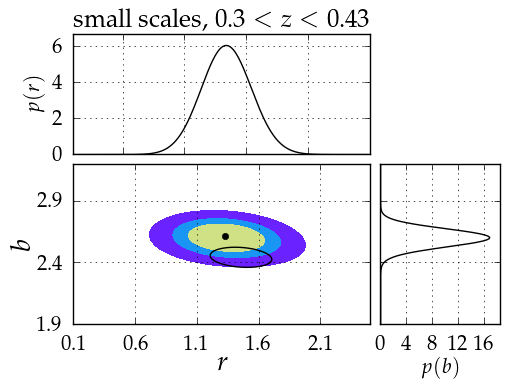}
 \caption{\changes{The $1$, $2$, and $3\sigma$ parameter constraints on the galaxy bias parameter 
          $b$ and $r$ for the LOWZ BOSS galaxy sample as well as the marginalized likelihoods of $b$ and $r$. 
          The black ellipse, if shown, is the 1$\sigma$ contour of the corresponding measurement using a 
          \textit{Planck} cosmology from the upper two panels. These parameters were constrained by 
          a maximum likelihood fit to the $\Upsilon_{\mathrm{gm}}$ and $\Upsilon_{\mathrm{gg}}$. All constraints 
          agree within $1\sigma$. 
          \textit{Top left:} Likelihood contours for the $3-20\, \rm{arcmin}$ interval using a 
          \textit{Planck} cosmology.
          \textit{Top right:} Likelihood contours for the $20-70 \, \rm{arcmin}$ interval using a 
          \textit{Planck} cosmology. 
          \textit{Middle left:} Likelihood contours for the $3-20 \, \rm{arcmin}$ interval using the \citet{heymans} 
          cosmology. 
          \textit{Middle right:} Likelihood contours for the $20-70 \, \rm{arcmin}$ interval using the \citet{heymans} 
          cosmology.
          \textit{Bottom left:} Likelihood contours for the $3-20 \, \rm{arcmin}$ interval using a \textit{Planck} cosmology 
          and the $0.15<z<0.3$ lens sample. 
          \textit{Bottom right:} Likelihood contours for the $3-20 \, \rm{arcmin}$ interval using a \textit{Planck} cosmology 
          and the $0.3<z<0.43$ lens sample.}}
 \label{fig:contours}
\end{figure*}

\changes{Furthermore, we conduct a redshift evolution test where we} split up the lens sample in two sub-samples with
$0.15<z<0.3$ and $0.3<z<0.43$. In this way we can test if the model is capable of describing these measurements in 
a proper way. We then make the same measurements as before using the \textit{Planck} cosmology and 
the $\vartheta\in [3',20']$ interval. This yields two new estimates for $b$ and for $r$. 
\changes{We find $b=2.35^{+0.04}_{-0.05}$ and $r=1.84^{+0.24}_{-0.23}$ for the low-redshift 
sample and $b=2.61^{+0.07}_{-0.08}$ and $r=1.33^{+0.21}_{-0.20}$ for
the high-redshift one. 
They are also shown in Table \ref{tab:res}}. The measured correlation functions are displayed
in Fig. \ref{fig:corr_fit} and the likelihood contours in
Fig. \ref{fig:contours}.  We find that $r$ becomes smaller for
the higher redshift sample\changes{, whereas $b$ gets larger. All estimates are, however, consistent 
with the parameters determined using the whole sample. In fact, the two sub-sample values for $b$ and $r$ 
bracket their whole sample counterparts.} 

\section{Discussion \& Outlook}\label{discussion}
We introduced a new estimator for galaxy-clustering, $\Upsilon_{\rm{gg}}$, and for galaxy-galaxy lensing, 
$\Upsilon_{\rm{gm}}$. Those are generalizations of the methods 
introduced and tested in \citet{baldauf} and \citet{mandelbaum}, respectively. The estimators are a 
discretisation of the $\Upsilon(\vartheta,\vartheta_{\rm min})$, which
leads to substantial data compression and thus a lower-dimensional covariance, while still
eliminating the sensitivity to the matter distribution on small scales.
Especially, \changes{lowering the dimension of the data covariance increases the accuracy in its measurement 
for a fixed number of mock realisations. Recall that} the number of 
mock realizations needed to find a good estimate of the covariance increases with the number of data points. 
We applied this method to data using the BOSS LOWZ sample as lenses and galaxies from the RCSLenS as sources. 
While fixing the cosmology, we performed a simultaneous fit to $\Upsilon_{\rm{gg}}$ and $\Upsilon_{\rm{gm}}$
with $b$ and $r$ as free parameters. For different angular scales as well as different assumed cosmologies we find 
$b$ slightly higher than the findings of \citet{boss_lowz} \changes{and \citet{chuang}. This tension could be resolved
if our assumption of scale-independent bias was a poor approximation to the true galaxy bias of this 
sample\changestwo{, as both of the studies mentioned allow for scale dependent bias.}}

On large angular scales, the cross-correlation coefficient $r$ is
found to be compatible with unity, as expected for the corresponding
spatial scales (e.g., B10). On the smaller angular
scale interval, we find a value for $r$ that is significantly larger
than unity, most likely due to a different scale and redshift dependence of
\changesfour{the various power spectra that enter the $\Upsilon$'s in Equations (\ref{eq:ups_gg_theory}) and
(\ref{eq:ups_gm_theory}), and our definition of the `effective' coefficients $b$ and $r$
as an average of the three-dimensional bias and correlation
coefficients $\hat b$ and $\hat r$.}
If one had already measured values for $b$ and $r$ this method can even be used for cosmological studies. 
In these studies it will be necessary to find out how many orders of $\Upsilon$ are sufficient to extract 
all cosmological information from the signal. As in this work it was not possible to do so as all information is 
already contained in the first few orders, due to our simplified bias
models. This might change in a cosmological analysis from
substantially larger data sets with more complex models.

Summarizing, the new estimators \changes{presented in this paper are
  promising tools for future} large-scale structure studies,
especially given their advantageous abilities concerning data
compression and the dimension of the data covariance.

\section*{Acknowledgements}
We are grateful to the RCS2 team for
planning the survey, applying for observing time, and conducting the
observations. We would like to thank Matthias Bartelmann for being our external
blinder, revealing which of the four catalogues analysed
was the true unblinded catalogue at the end of this study. 
We would like to thank Christopher Morrison and Patrick Simon for fruitful discussions. 
\changesfour{Also, we would like to thank our anonymous referee for constructive comments.}
We acknowledge use of the Canadian Astronomy Data Centre
operated by the Dominion Astrophysical Observatory for the National
Research Council of Canada's Herzberg Institute of Astrophysics. 
\changes{Computations for the N-body simulations were performed on the GPC 
supercomputer at the SciNet HPC Consortium. SciNet is funded by: the 
Canada Foundation for Innovation under the auspices of Compute Canada; 
the Government of Ontario; Ontario Research Fund - Research Excellence; 
and the University of Toronto.} 

AB was supported for this research partly through a stipend from the
International Max Planck Research School (IMPRS) for Astronomy and
Astrophysics at the Universities of Bonn and Cologne and through
funding from the Transregional Collaborative Research Centre `The dark
Universe' (TR 33) of the DFG.
\changesfour{PS acknowledges support from the DFG grant SCHN 342-13.}
\changes{HH is supported by an Emmy Noether grant (No. Hi 1495/2-1) of the Deutsche Forschungsgemeinschaft. 
CB acknowledges the support of the Australian Research Council through the award of a Future Fellowship. 
CH and LK acknowledge support from the European Research Council under FP7 grant number 240185. 
JHD is supported by the NSERC of Canada. 
RN acknowledges support from the German Federal Ministry for Economic Affairs and Energy (BMWi) provided 
via DLR under project no.50QE1103. 
MV acknowledges support from the European Research Council under FP7 grant number 279396 and the 
Netherlands Organisation for Scientific Research (NWO) through grants 614.001.103.}
For this work we made use of the correlation codes \texttt{athena}\footnote{\url{http://www.cosmostat.org/athena.html}} 
\citep{athena}. 
and \texttt{swot}\footnote{\url{https://github.com/jcoupon/swot}} \citep{coupon2}. 

\small
\changes{Author contributions:
All authors contributed to the development and writing of this paper. 
The authorship list reflects the lead authors of this paper (AB, PS and HH)
followed by \changestwo{two alphabetical groups. 
The first alphabetical group includes key contributers to the science analysis 
and interpretation in this paper, the founding core team and those whose
long-term significant effort produced the final RCSLenS data product. The
second group covers members of the RCSLenS team who made a significant 
contribution to the project, this paper, or both.}
HH led the RCSLenS collaboration.}

\bibliography{mylib}

\appendix

\section{Estimating $\Upsilon_{\mathrm{x}}$}
\begin{figure}
 \centering
 \includegraphics[width=8.5cm,height=8.5cm,keepaspectratio=true]{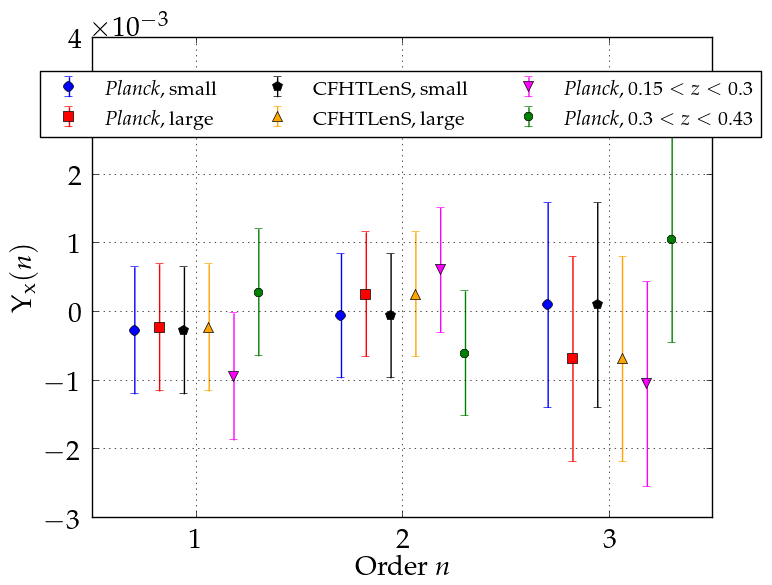}
 \caption{\changes{We present the $\Upsilon_{\mathrm{x}}$ for all six measurements conducted in this paper. 
          We find it to be always consistent with zero.}}
 \label{fig:ups_cross}
\end{figure}

\changes{For weak gravitational lensing measurements it is important to check if the cross shear, $\gamma_{\mathrm{x}}$
is consistent with zero. If not so, this points to systematic issues in the data. We can conduct a similar test for 
the $\Upsilon_{\mathrm{gm}}(n)$, where we replace $\gamma_{\mathrm{t}}$ with $\gamma_{\mathrm{x}}$ in 
Eq. (\ref{eq:ups_gm})
\begin{equation}
  \Upsilon_{\mathrm{x}}(n) = \int_{\vartheta_{\mathrm{min}}}^{\vartheta_{\mathrm{max}}} \mathrm d \vartheta \:
 \vartheta \: \: \mathcal{Q}_{n}(\vartheta) \: \gamma_{\mathrm{x}}(\vartheta) \;.
\end{equation}
As $\gamma_{\mathrm{x}}$ needs to be zero on all scales, so does its compressed counter part $\Upsilon_{\mathrm{x}}$. 
We estimated $\Upsilon_{\mathrm{x}}$ for all six measurements described in this paper and show the signal in 
Fig. \ref{fig:ups_cross}. Indeed, we find it to be consistent with zero for all three orders.}

\end{document}